\documentclass[a4paper,12pt,onecolumn,notitlepage]{article}
\usepackage[pdfpagemode=None,bookmarksopen=false,colorlinks=true,linkcolor=blue,citecolor=blue]{hyperref}
\usepackage{amsmath}
\usepackage{amssymb}
\usepackage{sectsty}
\usepackage{bibspacing}
\usepackage{rpkid}
\long\def\symbolfootnote[#1]#2{\begingroup%
\def\thefootnote{\fnsymbol{footnote}}\footnote[#1]{#2}\endgroup}
\allsectionsfont{\normalsize}
\begin{document}
\thispagestyle{empty}
\numberwithin{equation}{section}
\begin{center}
\noindent
{\Large \textbf{Information dynamics and new geometric foundations of quantum theory}}\symbolfootnote[1]{\scriptsize To be published in: Khrennikov A. (ed.), \textit{Proceedings of the Conference ``Foundations of Probability and Physics 6'', Linn\'{e}universitetet, V\"{a}xj\"{o}, June 13-16, 2011}, AIP Conf. Proc., Springer, Berlin.}\\
\ \\
{Ryszard Pawe{\l} Kostecki}\\
{\small\ \\}{\small
\textit{Institute of Theoretical Physics, University of Warsaw, Ho\.{z}a 69, 00-681 Warszawa, Poland}}{\small\\
\ \\
December 2, 2011}
\end{center}
\begin{abstract}{\small\noindent We discuss new approach to mathematical foundations of quantum theory, which is completely independent of Hilbert spaces and measure spaces. New kinematics is defined by non-linear geometry of spaces of integrals on abstract non-commutative algebras. New dynamics is defined by constrained maximisation of quantum relative entropy. We recover Hilbert space based approach (including unitary evolution and the von Neumann--L\"{u}ders rule) and measure theoretic approach to probability theory (including Bayes' rule) as special cases of our approach. } 
\end{abstract}
\section{Introduction}
Foundations of quantum theory consist of mathematical formalism, its conceptualisation, and particular methods of experimentation. The Hilbert space based mathematical formalism of quantum mechanics, proposed by Hilbert and developed by von Neumann \cite{Hilbert:vonNeumann:Nordheim:1927,vonNeumann:1932:grundlagen}, has a status of orthodoxy. However, it became also a subject of many critical assessments, including critique by von Neumann himself. As a result, several different alternative mathematical frameworks were developed, such as quantum logic approaches, algebraic approaches, convex set approaches, path integral approaches, Gel'fand triplet approaches, and semi-spectral approaches (to name only few). Yet, none of these frameworks was able to completely replace the orthodox setting. In our opinion, the Hilbert space based framework is relevant only for quantum \textit{mechanics}, which covers just a certain particular class of quantum \textit{theoretic} models, corresponding to a certain restricted class of experimental situations. There seems to be many experimental situations such that the corresponding quantum theoretic models do not fit into the orthodox framework without severe deformations. An important example is provided by quantum field theory, which fits into Hilbert space based frames only in perturbative sense, using many \textit{ad hoc} assumptions, and even some ill-founded techniques. This means that the mathematical formalism for quantum \textit{theory} in its full generality still waits to be developed. Such theory should include non-perturbative generic formulation of quantum field theory. The Hilbert space based framework for quantum \textit{mechanics} plays for us a role similar to \textit{special} relativity: it asks for a general theory with additional non-trivial geometric structures that would encode the `interaction'. In what follows we discuss the elements of a new approach to construction of such theory. For a full account of it, including its conceptualisation, see the forthcoming series of papers \cite{Kostecki:2011:towards:series}.
\section{New foundations}
Given abstract non-commutative $W^*$-algebra $\N$, we define the \textit{quantum information model} $\M(\N)$ as a subset of the space $\N_*^+$ of normal positive finite linear $\CC$-valued functions on $\N$. A \textit{predual} $\N_*$ is defined as such Banach space that $\N$ is its Banach dual: $(\N_*)^\banach\iso\N$. The spaces $\M(\N)$ can be equipped with several different geometric structures, including the structures of differential and convex geometry. Any particular collection of geometric structures on $\M(\N)$ will be called its \textit{quantum information geometry}. We propose to consider the space $\M(\N)$ together with its (non-linear) quantum information geometry as a replacement for linear Hilbert space $\H$ (and linear operators over it) in the role of kinematic setting of a quantum theory.

This proposition is motivated by reconsideration of quantum theory as a replacement of probability theory (and not of classical mechanics). The spaces $\M(\N)$ are just non-normalised non-commutative analogues of probabilistic models 
\[
        \M(\X,\mho(\X),\mu)\subseteq L_1(\X,\mho(\X),\mu)^+,
\]
where $\X$ is a sample space, $\mho(\X)$ is a boolean algebra of some countably additive subsets of $\X$, and $\mu$ is a probability measure on $\mho(\X)$. This analogy can be made precise by means of the Falcone--Takesaki non-commutative integration theory \cite{Falcone:Takesaki:2001} and abstract theory of integration over boolean algebras \cite{Fremlin:2000}, which allow to separate the essential elements of mathematical formalism of probability theory and quantum theory from the representation-dependent elements.

In particular, there are many different measure spaces $(\X,\mho(\X),\mu)$ which might not be related by any transformation or isomorphism, but lead anyway to isometrically isomorphic commutative $L_1$ spaces $L_1(\X,\mho(\X),\mu)$. This can be made precise in the following way. Define \textit{camDcb-algebra} as a countably additive, Dedekind complete, boolean algebra allowing a semi-finite strictly positive measure. Every localisable measure space $(\X,\mho(\X),\mu)$ allows to construct a corresponding camDcb-algebra $\mho$ by 
\[
        \mho:=\mho(\X)/\{x\in\mho(\X)\mid\mu(x)=0\}.
\]
(Localisability is equivalent with validity of the Steinhaus--Nikod\'{y}m duality 
\[
        L_1(\X,\mho(\X),\mu)^\banach\iso L_\infty(\X,\mho(\X),\mu),
\]
which is necessary to consider elements of probabilistic model as functions given by the Radon--Nikod\'{y}m derivatives.) Moreover, to each camDcb-algebra $\mho$ there can be associated a range of commutative $L_p$ spaces $L_p(\mho)$, $p\in[1,\infty]$, which are independent of the choice of measure on $\mho$. (They are defined by an abstract integration theory on a Riesz lattice $\stone(\mho)$ of characteristic functions on the Stone spectrum of $\mho$.) The association of $L_p(\mho)$ is functorial over a category of camDcb-algebras with isomorphisms as arrows. Thus, all isometrically isomorphic $L_1(\X,\mho(\X),\mu)$ spaces are just representations of a single $L_1(\mho)$ space, associated with a particular camDcb-algebra $\mho$. For any given camDcb-algebra $\mho$ and some integral $\omega$ on $\stone(\mho)$, the Loomis--Sikorski theorem provides a representation of $\mho$ in terms $(\X,\mho(\X),\mu)$. Thus, probabilistic models can be defined independently of the representation in terms of measure spaces, as subsets of normalised part of $L_1(\mho)^+$ for a given commutative camDcb-algebra $\mho$.

The Falcone--Takesaki theory provides a generalisation of the representation-independent construction of the spaces of integrals to the case when underlying integrable algebra is non-commutative (i.e., it is a $W^*$-algebra). It includes even the functorial association of a range of non-commutative $L_p$ spaces $L_p(\N)$, $p\in[1,\infty]$, which are independent of the choice of weight on $\N$. The isometric isomorphisms 
\[
        (\N_*)^\banach\iso L_1(\N)^\banach\iso L_\infty(\N)\iso\N
\]
correspond to 
\[
        L_1(\mho)^\banach\iso L_\infty(\mho).
\]
Moreover, for any pair of a $W^*$-algebra $\N$ and an integral $\omega\in\N_*^+$ the Gel'fand--Na\u{\i}mark--Segal (GNS) theorem associates a unique Hilbert space $\H_\omega$ and a unique (up to unitary isomorphism) representation $\pi_\omega:\N\ra\BBB(\H_\omega)$ such that there exists $\Omega_\omega\in\H_\omega$ that is cyclic for $\pi_\omega(\N)$, and 
\[
        \omega(x)=\s{\Omega_\omega,\pi_\omega(x)\Omega_\omega}_\omega\;\;\;\forall x\in\N.
\]
(If $\N$ contains no type III factor and if $\N_*^+$ contains at least one faithful element $\omega$, then $\M(\N)$ can be represented as a space $\M(\H_\omega)$ of \textit{non-normalised} density operators over $\H_\omega$. However, type III factors arise naturally in quantum field theory, so this representation has a restricted validity.) Finally, for every commutative $W^*$-algebra $\N$ there exists such camDcb-algebra $\mho$ that $\N\iso L_\infty(\mho)$. This turns $\M(\N)\subseteq\N_*^+$ to $\M(\mho)\subseteq L_1(\mho)^+$. In face of such deep structural relationships between commutative and non-commutative integration theory it is quite hard to find convincing arguments in favour of consideration of probability theory and quantum theory as two separate theories. This way probability theory becomes precisely a special (commutative and normalised) case of quantum theory, at least at the level of kinematics of both theories. We do not require normalisation of $\M(\mho)$ or $\M(\N)$, because it is required only by the relative frequency interpretation, which: (1) has failed to provide sound conceptual and mathematical foundations of probability theory; (2) is impossible in quantum case, because integrals on $W^*$-algebras (beyond type II$_1$ factors) do not allow representation in form of relative frequency \cite{vonNeumann:1937,Redei:1999}.

It follows that von Neumann's spectral representation theorem, which represents Hilbert space based quantum theoretic model in terms of measure space based  probabilistic model (expressed in terms of commutative $L_2(\X,\mho(\X),\mu)$ space) is just a translation from some very particular representation of quantum model $\M(\N)$ to some very particular representation of probabilistic model $\M(\mho)$. Thus, neither this theorem nor the eigenvalues of elements of some particular Hilbert space representation of $\N$ would play foundational role in our approach to quantum theory. (On the other hand, the non-normalised expectation values play the role of parameters on $\M(\N)$.) More generally, we do not see any convincing arguments requiring to analyse general models $\M(\N)$ in terms of some special models $\M(\mho)$: quantum theoretic models should be quantitatively constructed and analysed on their own right, without passing to commutative normalised sector (probability theory). To achieve this goal, we consider quantum kinematics as given by models $\M(\N)$ \textit{together with} the quantum information geometric structures on $\M(\N)$, and require that the latter should be directly related with (operational) experimental descriptions, without invoking probability theory. In this sense, we postulate that the usual approach to prediction of experimental behaviour based on ``quantum mechanics + methods of quantum model construction + probability theory + methods of statistical inference'' should be replaced by ``quantum theory + methods of construction of kinematics and dynamics of quantum models''.
\section{Quantum geometry\label{quantum.geometry.section}}
Among different geometric structures on $\M(\N)$, a particularly important role is played by the non-symmetric distance function, 
\[
        \M(\N)\times\M(\N)\ni(\phi,\omega)\mapsto D(\phi,\omega)\in[0,+\infty]
\]
such that $D(\phi,\omega)=0\iff\phi=\omega$, called \textit{quantum information deviation} or negative \textit{quantum relative entropy}. It allows to determine several other geometric objects on $\M(\N)$, see e.g. \cite{Hasegawa:1993,Lesniewski:Ruskai:1999,Jencova:2004:entropies}. In particular, if 
\[
        \M(\N)\subseteq\N_{*01}^+:=\{\omega\in\N_*^+\mid\omega(\II)=1,\;\omega(x^*x)=0\limp x=0\},
\]
then $\M(\N)$ can be equipped with the structure of differential manifold defined in terms of quantum relative entropic perturbations \cite{Jencova:2010}. In such case, if $\dim\M(\N)<\infty$ and $D$ has a non-negative hessian, then the riemannian metric $g$ and a pair $(\nabla,\nabla^\nsdual)$ of affine connections on $\M(\N)$ can be defined by \begin{align*}
        g_\phi(u,v)&:=(\partial_u)_\phi(\partial_v)_\omega D(\phi,\omega)|_{\omega=\phi},\\
        g_\phi((\nabla_u)_\phi v,w)&:=-(\partial_u)_\phi(\partial_v)_\phi(\partial_w)_\omega D(\phi,\omega)|_{\omega=\phi},\\
        g_\phi(v,(\nabla^\nsdual_u)_\phi w)&:=-(\partial_u)_\omega(\partial_w)_\omega(\partial_v)_\phi D(\phi,\omega)|_{\omega=\phi},
\end{align*}
where $(\partial_u)_\phi$ is a directional derivative at $\phi\in\M(\N)$ in the direction $u\in\T_\phi\M(\N)$. Here $\T_\phi\M(\N)$ can be identified with any of the sets 
\[
        \{x\varphi^\gamma\in L_{1/\gamma}(\N)\mid\re\phi([\DD\phi:\DD\varphi]_{i\gamma}(x))=0\},
\]
for $\gamma\in]0,1]$, where $[\DD\phi:\DD\varphi]_z$ is Connes' derivative (the non-commutative Radon--Nikod\'{y}m derivative). Every such triple $(g,\nabla,\nabla^\nsdual)$ forms a Norden--Sen geometry on $\M(\N)$, characterised by the equation 
\[
        g(\nabla_uv,w)+g(v,\nabla^\nsdual_uw)=u(g(v,w))\;\;\forall u,v\in\T\M(\N).
\]
It naturally determines an associated riemannian geometry $(g,\bar{\nabla})$ on $\M(\N)$, with
\[
        \bar{\nabla}:=\frac{1}{2}(\nabla+\nabla^\nsdual).
\]
If $D_\Psi$ is a quantum Bregman deviation \cite{Petz:2007:Bregman,Kostecki:2011:OSID}, defined using the embedding $\ell_\Psi:\M(\N)\ra L$ into a complex linear space $L$ and a convex function $\Psi:L\ra]-\infty,+\infty]$, and if, given $\Q\subseteq\M(\N)$, $\ell_\Psi(\Q)\subseteq L$ is non-empty, closed and convex, then the non-linear \textit{Bregman projection} 
\[
        \PPP^{\Psi}_{\Q}:\M(\N)\ra\Q
\]
is defined by the condition 
\[
        \forall\omega\in\M(\N)\;\;\exists!\PPP^{\Psi}_\Q(\omega)=\arg\inf_{\phi\in\Q}\{D_\Psi(\phi,\omega)\}.
\]
It satisfies the \textit{generalised pythagorean equation}: 
\[
        D_\Psi(\varphi,\PPP^{\Psi}_\Q(\phi))+D_\Psi(\PPP^{\Psi}_\Q(\phi),\omega)=D_\Psi(\varphi,\omega)\;\;\forall\varphi\in\Q\;\;\forall\omega\in\M(\N).
\]
The Hilbert space norm distance, $\n{\xi_1-\xi_2}_\H^2$ is (up to a factor of $\frac{1}{2}$) a special case of Bregman deviation $D_\Psi(\phi_1,\phi_2)$ for linear space $L$ given by $L_2(\N)\iso\H$, $\Psi(\cdot)=\n{\cdot}^2$, and $\xi(\phi):=\ell_\Psi(\phi)=2\phi^{1/2}$. In such case $\PPP^{\Psi}_\Q$ are norm-orthogonal projections. Hence, \textit{quantum information geometry of $\M(\N)$ reduces in special cases to riemannian geometry and to the complex Hilbert space geometry}.
\section{Quantum dynamics}
Under our particular concern is a family of quantum relative entropies that belong to the class of Bregman deviations and are non-increasing under the completely positive trace-preserving maps $T^\coa$, 
\[
        D(\omega,\phi)\geq D(T^\coa(\omega),T^\coa(\phi)).
\]
In \cite{Kostecki:2011:OSID} we defined the family of quantum deviations 
\[
        D_\gamma(\omega,\phi):=\int(\omega/(1-\gamma)+\phi/\gamma-\re(\omega^\gamma\phi^{1-\gamma})/(\gamma(1-\gamma)))
\]
for $\gamma\in]0,1[$ (and by its limits under integral for $\gamma\in\{0,1\}$) and showed that it satisfies both above conditions. We have conjectured that it is characterised by these conditions. This provides a strong restriction on the representations of models $\M(\N)$ in terms of linear spaces $L$. The Bregman deviations $D_\gamma(\omega,\phi)$ are based on the embeddings 
\[
        \ell_\gamma:\M(\N)\ni\phi\mapsto\frac{1}{\gamma}\phi^\gamma\in L_{1/\gamma}(\N),
\]
restricting $L$ to one of $L_{1/\gamma}(\N)$ spaces. The differentiation of $D_\gamma$ generates the Wigner--Yanase--Dyson metrics for $\gamma\in]0,1[$ and the Bogolyubov--Kubo--Mori metrics for $\gamma\in\{0,1\}$. 

We define the \textit{quantum information dynamics} as a mapping provided by the variational principle of constrained maximisation of quantum relative entropy 
\[
        \M(\N)\ni\omega\mapsto\PPP^{\gamma}_\Q(\omega):=\arg\inf_{\phi\in\M(\N)}\{D_\gamma(\omega,\phi)+F(\phi)\}\subset\M(\N),
\]
where $F:\Q\ra]-\infty,\infty]$ represents the constraints \cite{Kostecki:2010:AIP}. (More generally, one might replace $D_\gamma(\omega,\phi)$ by 
\[
        \int_{\varphi\in\M(\N)}E(\varphi,\omega)D_\gamma(\varphi,\phi),
\]
where $E(\cdot,\omega)$ is a positive measure on $\M(\N)$ and define the constraints in terms of $F$ \textit{and} $E$.) The solution to this dynamics exists and is unique if $F$ is weakly lower semi-continuous, convex, $F\not\equiv+\infty$, and if $\ell_{1-\gamma}(\Q)$ is non-empty, weakly closed and convex subset of $L_{1/(1-\gamma)}(\N)$ space. We refer to this mapping as `information dynamics', because it quantifies the changes of knowledge, which are imposed by the constraints (evidence) $F$. The constraints $F$ depend not only on $\phi$, but also on some function $\operationalf(\theta_1,\ldots,\theta_n)$ of operational `experimental evidence' $(\theta_1,\ldots,\theta_n)$. A simple example is
\begin{equation}
        F(\phi,\theta_1,\ldots,\theta_n)=\lambda_1(\phi(\II)-1)+\lambda_2(\phi(x)-\operationalf(\theta_1,\ldots,\theta_n)),
\label{constraints}
\end{equation}
where $x\in\N$ is some abstract quality (e.g. `energy'), while $\lambda_1$ and $\lambda_2$ are Lagrange multipliers. The constraint \eqref{constraints} \textit{defines} the knowledge about the particular quantification of the abstract quality $x$ (here given by an expectation $\phi(x)$) that \textit{is considered} to correspond to the given function $\operationalf$ of given `experimental evidence' $(\theta_1,\ldots,\theta_n)$. Note that this allows for consideration of $\operationalf(\theta_1,\ldots,\theta_n)=\frac{1}{n}\sum_{i=1}^n\theta_i$, with $\theta_i$ defined as a particular value attained at the registration scale of some `energy-measuring device', as well as of $\operationalf(\theta_1,\ldots,\theta_n)=\theta_1$, with $\theta_1$ defined as an arithmetic average of these values. The choice among various possibilities of such kind determines the particular operational meaning of the results of of quantum information dynamics. (In general, we consider references to classical mechanics or probability theory as irrelevant to mathematical foundations and experimental verification of quantum theory. The experimental verification of quantum theoretic models requires only some language for operational description of experimental situations and some rules relating operational descriptions of particular experiments with corresponding quantum theoretical models. For more discussion of interpretation of our mathematical framework, see \cite{Kostecki:2010:AIP,Kostecki:2011:principles,Kostecki:2011:towards:series}.)

If the constraints $F$ are parametrised by some $t$ interpreted as `time', $F(\phi)=F(\phi,t)$ with $F(\phi,t=t_0)=0$, then the trajectory 
\[
        t\mapsto\omega(t):=\PPP^{\gamma}_{F(t)}(\omega(t=t_0))
\]
can be understood as a non-linear quantum \textit{temporal evolution} on $\M(\N)$ (the domain of $t$ might be continuous or discrete).
\section{Recovery of orthodox formalisms}
The measure space based framework for probability theory is recovered by passing to camDcb-algebra $\mho$ via $L_\infty(\mho)\iso\N$ for commutative $\N$, and by the Loomis--Sikorski measure space representation of $\mho$. The models $\M(\N)$ turn then into $\M(\X,\mho(\X),\mu)$, while their quantum information geometry and quantum information dynamics turn to their commutative counterparts on $\M(\X,\mho(\X),\mu)$. The normalisation condition imposes the projection of geometry on unit sphere in $L_1(\mho)$ and the additional constraint on information dynamics. Finally, for $\gamma=0$, $\dim\M(\mho)<\infty$ with $\theta:\M(\mho)\ra\RR^n$, and $F$ represented as 
\[
        F(q)=\lambda_1(\int dx\int d\theta q(x|\theta)-1)+\lambda_2(\int d\theta q(x|\theta)-\delta(x-b))
\]
the information dynamics $\omega\mapsto\PPP^{\gamma}_\Q(\omega)$ reduces to Bayes' rule 
\[
        p(x|\theta)\mapsto p(x|\theta)p(b|x\land\theta)/p(b|\theta),
\]
see \cite{Caticha:Giffin:2006}. This way probability theory is a special case of quantum theory also on the level of dynamics of both theories.

The Hilbert space based kinematic framework for quantum theory can be reconstructed as a particular case of quantum geometric approach. The $L_2(\N)$ space can be naturally equipped with the scalar product defined by 
\[
        \s{x,y}:=\int y^*x,
\]
and it is complete in the norm topology generated by this product. As a Hilbert space, $L_2(\N)$ is unitary isomorphic to the Hilbert space $\H_H$ of Haagerup's standard representation $\pi_H:\N\ra\BBB(\H_H)$. In consequence, $\M(\N)$ can be represented using 
\[
        \ell_{1/2}:\M(\N)\ni\phi\mapsto\ell_{1/2}(\phi)=2\phi^{1/2}\in L_2(\N)\iso\H_H.
\]
Moreover, for any faithful $\omega\in\N_*^+$, the Haagerup representation is unitary equivalent with the GNS representation $\pi_\omega:\N\ra\BBB(\H_\omega)$. As explained in Section \ref{quantum.geometry.section}, the Bregman projections of $D_{1/2}$ on $\M(\N)$ (naturally associated with $\ell_{1/2}$) turn into norm-orthogonal projections on $\H_H$. This allows to consider orthodox quantum theoretic kinematics as a linear representation of a special case of quantum information geometric kinematics.

In the Hilbert space based framework there are two different notions of temporal evolution: the unitary evolution and the non-unitary evolution. The latter changes the probabilistic predictions (inferences) drawn from the formalism, while the former does not change them. Hence, we consider the unitary evolution as a part of information kinematics, as opposed to non-unitary evolution which belongs to information dynamics. The von Neumann--L\"{u}ders rule 
\[
        \M(\H)\ni\rho\mapsto\sum_iP_i\rho P_i\in\M(\H),
\]
with $\{P_i\}$ given by the projection operators arising from orthogonal decomposition of unit $\II\in\BH$ and such that $[P_i,\rho]=0$, is often considered as a non-commutative analogue of Bayes' rule (see e.g. \cite{Bub:1977,Schack:Brun:Caves:2001,Fuchs:2002}). In \cite{Kostecki:2010:Torun} we have conjectured that this analogy can be made strict in the following sense: \textit{the von Neumann--L\"{u}ders rule is a special case of constrained quantum relative entropy maximisation}. This conjecture has been proved in \cite{GHKK:2011} for $\dim\M(\N)<\infty$ and $\gamma=0$, which corresponds to the Umegaki deviation
\[
        D_0(\omega,\phi)=\tr(\rho_\phi(\log\rho_\phi-\rho_\omega)),     
\]
and for $\Q=\{\sum_iP_i\rho_i P_i,\;\rho_i\geq0\}$, which is convex and closed in terms of the coordinate embedding 
\[
        \ell_0:\M(\N)\ni\rho\mapsto\log\rho\in L_\infty(\N)\iso\N.
\]
This result allows to replace \textit{ad hoc} von Neumann--L\"{u}ders updating of quantum state of knowledge (which restricts the allowed `experimental evidence' to projections in commutative $L_2$ space) by the general principle of quantum information dynamics with constraints defined by convex closed subsets of non-commutative $L_p(\N)$ spaces. As a result, our approach allows for more flexible operational specification of the `experimental evidence', and for deriving various `quantum measurement' rule from a single underlying principle. 

The models $\M(\N)$ allow different temporal trajectories generated by the entropic dynamics, so their points correspond in general to quantum information states in various instances of time $t$. Because orthodox approach defines quantum models in a single instant of time $t$, $\M(\N)$ should allow for a global foliation by hypersurfaces of codimension 1. These hypersurfaces will be indexed by $s\in\RR$. The parametrisation $s$ is a priori independent of time $t$, except of the condition that $t$ has to be equipped with a partial order relation, and there must be given an order isomorphism between the partial order relations of $t$ and $s$. For each leaf $\M_s(\N)$ there should exist an associated faithful reference state $\phi(s)\in\N_*^+$, allowing to consider 
\[
        \ell_{1/2}:\M_s(\N)\ra\H_{\phi(s)}. 
\]
In particular, the family $\phi(s)$ can be given by a continuous trajectory $\phi(s)\in\M(\N)$ such that $\phi(s)\in\M_s(\N)$.

If some $\H_{\phi(s)}$, say $\H_{\phi(0)}$, is fixed as \textit{the} orthodox Hilbert space $\H$, then the unitary equivalence with other $\H_{\phi(s)}$ becomes represented in terms of a family $U(s)$ of unitary operators on $\H$. If $U(s)$ is a group, then it becomes identified with the orthodox unitary evolution. Thus, the construction of unitary evolution has information kinematic character and amounts to a choice of a global foliation of $\M(\N)$ and a choice of a family $\phi(s)$ of reference states associated to it. The hamiltonians constructed in this way are of purely epistemic character. 

According to the Tomita--Takesaki theorem, each faithful element $\omega$ of $\N_*^+$ determines a unique unitary automorphism $\RR\ni r\mapsto\sigma^\omega_r\in\Aut(\pi_\omega(\N))$ on $\H_\omega$ such that $\omega\circ\sigma^\omega=\omega$. This leads to as whether $U(s)$ could be related with $\sigma^\omega_r$, at least in a special case them $\phi(s)=\phi(0)\;\forall s\in\RR$. The automorphism $\sigma^{\phi(0)}_r$ has a virtue of being canonically associated to each faithful $\phi(0)$, but has also a drawback of excluding various quantum hamiltonians that do not agree with the properties of the hamiltonian $\sigma^{\phi(0)}_r$. We will discuss this problem elsewhere.

The two temporal evolutions of Hilbert space based setting are dependent on two different time parameters, respectively: `external time' $t$ of the constraints $F(\phi,t)$ and the `internal time' $s\in\RR$ of the global foliation of $\M(\N)$. This is not in conflict with the orthodox approach, because the latter simply avoids answering the question about temporal relationship of `time of measurement' and `time of unitary evolution'. The relation between $s$ and $t$ will be discussed in more details in \cite{Kostecki:2011:towards:series} and is directly related with reconsideration of models $\M(\N)$ as \textit{quantum space-times} \cite{Duch:Kostecki:2011}.
\section*{Acknowledgments}
{\small I am indebted to F.~Hellmann, C.J.~Isham, W.~Kami\'{n}ski, and S.L.~Woronowicz for valuable discussions. Support by FNP \textit{Mistrz} 2007,  ESF \textit{QGQG} 2706, MSWiN \textit{182/N} QGG/2008/0 and NCN \textit{N N202} 343640 grants is acknowledged.}
{\scriptsize
\bibliographystyle{../rpkbib}
\bibliography{../rpkrefs}

\begin{thebibliography}{10}

\bibitem{Bub:1977}
{Bub J.}, 1977, \textit{Von Neumann's projection postulate as a probability
  conditionalization rule in quantum mechanics}, J. Phil. Logic \textbf{6},
  381.

\bibitem{Caticha:Giffin:2006}
{Caticha A., Giffin A.}, 2006, \textit{Updating probabilities}, in:
  Mohammad-Djafari A. (ed.), \textit{Bayesian inference and maximum entropy
  methods in science and engineering}, AIP Conf. Proc. \textbf{872}, 31.
  \href{http://www.arxiv.org/pdf/physics/0608185}{arXiv:physics/0608185}.

\bibitem{Duch:Kostecki:2011}
{Duch P., Kostecki R.P.}, 2011, \textit{Quantum Schwarzschild space-time},
  submitted to Class. Quant. Grav.
  \href{http://www.arxiv.org/pdf/1110.6566}{arXiv:1110.6566}.

\bibitem{Falcone:Takesaki:2001}
{Falcone T., Takesaki M.}, 2001, \textit{The non-commutative flow of weights on
  a von Neumann algebra}, J. Funct. Anal. \textbf{182}, 170. Available at:
  \href{http://www.math.ucla.edu/~mt/papers/QFlow-Final.tex.pdf}{www.math.ucla.edu/$\sim$mt/papers/QFlow-Final.tex.pdf}.

\bibitem{Fremlin:2000}
{Fremlin D.H.}, 2000, 2001, 2002, 2003, \textit{Measure theory}, Vol.1-4,
  Torres Fremlin, Colchester.

\bibitem{Fuchs:2002}
{Fuchs C.A.}, 2002, \textit{Quantum mechanics as quantum information (and only
  a little more)},
  \href{http://www.arxiv.org/pdf/quant-ph/0205039}{arXiv:quant-ph/0205039}.

\bibitem{GHKK:2011}
{Guedes C., Hellmann F., Kami\'{n}ski W., Kostecki R.P.}, 2011,
  \textit{Entropic dynamics in quantum mechanics}, in preparation.

\bibitem{Hasegawa:1993}
{Hasegawa H.}, 1993, \textit{$\alpha$-divergence of the non-commutative
  information geometry}, Rep. Math. Phys. \textbf{33}, 87.

\bibitem{Hilbert:vonNeumann:Nordheim:1927}
{Hilbert D., von Neumann J., Nordheim L.}, 1927, \textit{\"{U}ber die
  Grundlagen der Quantenmechanik}, Math. Ann. \textbf{98}, 1.

\bibitem{Jencova:2004:entropies}
{Jen\v{c}ov\'{a} A.}, 2004, \textit{Generalized relative entropies as contrast
  functionals on density matrices}, Int. J. Theor. Phys. \textbf{43}, 1635.
  Available at:
  \href{http://www.mat.savba.sk/~jencova/ijtp.pdf}{www.mat.savba.sk/$\sim$jencova/ijtp.pdf}.

\bibitem{Jencova:2010}
{Jen\v{c}ov\'{a} A.}, 2010, \textit{On quantum information manifolds}, in:
  Gibilisco P. et al (eds.), \textit{Algebraic and geometric methods in
  statistics}, Cambridge University Press, Cambridge.

\bibitem{Kostecki:2010:Torun}
{Kostecki R.P.}, 2010, \textit{Information dynamics and geometric foundations
  of quantum theory}, talk given at \textit{42nd Symposium on Mathematical
  Physics}, Toru\'{n}, Poland.

\bibitem{Kostecki:2010:AIP}
{Kostecki R.P.}, 2010, \textit{Quantum theory as inductive inference}, in:
  Mohammad-Djafari A., Bercher J., Bessi\`{e}re P. (eds.), \textit{Proceedings
  of the 30th International Workshop on Bayesian Inference and Maximum Entropy
  Methods in Science and Engineering}, AIP Conf. Proc. \textbf{1305}, Springer,
  Berlin, p.24. \href{http://www.arxiv.org/pdf/1009.2423}{arXiv:1009.2423}.

\bibitem{Kostecki:2011:OSID}
{Kostecki R.P.}, 2011, \textit{The general form of $\gamma$-family of quantum
  relative entropies}, Open Sys. Inf. Dyn. \textbf{18}, 191.
  \href{http://www.arxiv.org/pdf/1106.2225}{arXiv:1106.2225}.

\bibitem{Kostecki:2011:principles}
{Kostecki R.P.}, 2011, \textit{On principles of inductive inference}, to be
  published in: Goyal P. (ed.), \textit{Proceedings of 31st International
  Workshop on Bayesian Inference and Maximum Entropy Methods in Science and
  Engineering, 10-15 July 2011, Waterloo}, AIP Conf. Proc., Springer, Berlin.
  \href{http://www.arxiv.org/pdf/1109.3142}{arXiv:1109.3142}.

\bibitem{Kostecki:2011:towards:series}
{Kostecki R.P.}, 2011, \textit{Towards new foundations of quantum theory I, II,
  III}, to be submitted.

\bibitem{Lesniewski:Ruskai:1999}
{Lesniewski A., Ruskai M.B.}, 1999, \textit{Monotone riemannian metrics and
  relative entropy on non-commutative probability space}, J. Math. Phys.
  \textbf{40}, 5702.
  \href{http://arxiv.org/PS_cache/math-ph/pdf/9808/9808016v1.pdf}{
  arXiv:math-ph/9808016}.

\bibitem{Petz:2007:Bregman}
{Petz D.}, 2007, \textit{Bregman divergence as relative operator entropy}, Acta
  Math. Hungar. \textbf{116}, 127. Available at:
  \href{http://www.renyi.hu/~petz/pdf/112bregman.pdf}{www.renyi.hu/$\sim$petz/pdf/112bregman.pdf}.

\bibitem{Redei:1999}
{R\'{e}dei M.}, 1999, \textit{``Unsolved problems in mathematics'': J. von
  Neumann's address to the International Congress of Mathematicians, Amsterdam,
  September 2--9, 1954}, Math. Int. \textbf{21}, 7. Available at:
  \href{http://phil.elte.hu/~redei/cikkek/intel.pdf}{phil.elte.hu/$\sim$redei/cikkek/intel.pdf}.

\bibitem{Schack:Brun:Caves:2001}
{Schack R., Brun T.A., Caves C.M.}, 2001, \textit{Quantum Bayes rule}, Phys.
  Rev. A \textbf{64}, 014305.
  \href{http://www.arxiv.org/pdf/quant-ph/0008113}{arXiv:quant-ph/0008113}.

\bibitem{vonNeumann:1932:grundlagen}
{von Neumann J.}, 1932, \textit{Mathematische Grundlagen der Quantenmechanik},
  Springer, Berlin. (engl. transl. 1955, \textit{Mathematical foundations of
  quantum mechanics}, Princeton University Press, Princeton).

\bibitem{vonNeumann:1937}
{von Neumann J.}, 1937, \textit{Quantum logic (strict- and probability
  logics)}, unpublished manuscript, John von Neumann Archive, Library of
  Congress, Washington (reviewed by Taub A.H., 1961, in: von Neumann J., 1961,
  \textit{Collected Works}, Pergamon, London, p.195).

\end{thebibliography}
}
\end{document}